\documentclass[10pt,a4paper,twoside]{article}

\newif\ifpgfplots
\pgfplotstrue

\usepackage[utf8]{inputenc}
\usepackage{latexsym,amssymb}
\usepackage{enumitem}
\setlist{itemsep=0ex,parsep=0ex,topsep=1ex}
\usepackage{ebproof}
\usepackage{coqdoc}
\usepackage{amsmath}
\usepackage{listings}
\usepackage{pgfplots}
\usepackage{pgfplotstable}
\pgfplotsset{compat=1.7}

\usepackage{subcaption}
\usepackage{graphicx}
\pgfplotsset{width=10cm,compat=1.9}

\newcommand{\ose}{\Downarrow}
\newcommand{\pose}{\Downarrow_p}

\newcommand{\bosse}[5]{\langle #1, #2, #3 \rangle \ose \{ #4, #5 \}}
\newcommand{\pbosse}[5]{\langle #1, #2, #3 \rangle \pose \{ #4, #5 \}}

\newcommand{\inlcv}[1]{\coqdocvar{inl}\ \coqdocvar{#1}}
\newcommand{\inrcv}[1]{\coqdocvar{inr}\ \coqdocvar{#1}}
\newcommand{\inl}[1]{\coqdocvar{inl}\ #1}
\newcommand{\inr}[1]{\coqdocvar{inr}\ #1}

\newtheorem{theo}{Theorem}

\newtheorem{hypo}[theo]{Hypothesis}

\pgfplotsset{compat=newest}
\usetikzlibrary{decorations.pathmorphing}

\pgfplotsset{
	SmallBarPlot/.style={
		font=\footnotesize,
		ybar,
    axis lines*=left,
		width=\linewidth,
		ymin=0,
		xtick=data,
		xticklabel style={text width=1.5cm, align=center},
		legend cell align={left},
	},
	BlueBars/.style={
		fill=blue!20, bar width=8pt
	},
	RedBars/.style={
		fill=red!20, bar width=8pt
	},
	GreenBars/.style={
		fill=green!20, bar width=8pt
	},
	BlueBars2/.style={
		fill=blue!20, bar width=16pt
	},
	RedBars2/.style={
		fill=red!20, bar width=16pt
	},
	GreenBars2/.style={
		fill=green!20, bar width=16pt
	},
  BlueBars3/.style={
		fill=blue!20, bar width=10pt
	},
	RedBars3/.style={
		fill=red!20, bar width=10pt
	},
	GreenBars3/.style={
		fill=green!20, bar width=10pt
	}
}
\pgfplotsset{select coords between index/.style 2 args={
		x filter/.code={
			\ifnum\coordindex<#1\fi
			\ifnum\coordindex>#2\fi
		}
}}

\begin{document}

\begin{center}{\Large\bf
A Comparison of Big-step Semantics Definition Styles
}\end{center}
\begin{center}
{\large\bf\noindent
Péter Bereczky$^1$, %
Dániel Horpácsi$^1$, %
Simon Thompson$^{1,2}$
}
\\[2mm]
ELTE Eötvös Loránd University, Department of Programming Languages and Compilers$^1$ \\
University of Kent, School of Computing$^2$
\\[1mm]
\texttt{
berpeti@inf.elte.hu, %
daniel-h@elte.hu, %
S.J.Thompson@kent.ac.uk
}
\end{center}
\vspace*{7mm}

\begin{abstract}

Formal semantics provides rigorous, mathematically precise definitions of programming languages, with which we can argue about program behaviour and program equivalence by formal means; in particular, we can describe and verify our arguments with a proof assistant. 
There are various approaches to giving formal semantics to programming languages, at different abstraction levels and applying different mathematical machinery: the reason for using the semantics determines which approach to choose. 

In this paper we investigate some of the approaches that share their roots with  traditional relational big-step semantics, such as (a) functional big-step semantics (or, equivalently, a definitional interpreter), (b) pretty-big-step semantics and (c) traditional natural semantics. We compare these approaches with respect to the following criteria: executability of the semantics definition, proof complexity for typical properties (e.g. determinism) and the conciseness of expression equivalence proofs in that approach. We also briefly discuss the complexity of these definitions and the coinductive big-step semantics, which enables reasoning about divergence.

To enable the comparison in practice, we present an example language for comparing the semantics: a sequential subset of Core Erlang, a functional programming language, which is used in the intermediate steps of the Erlang/OTP compiler. We have already defined a relational big-step semantics for this language that includes treatment of exceptions and side effects. The aim of this current work is to compare our big-step definition for this language with a variety of other equivalent semantics in different styles from the point of view of testing and verifying code refactorings.

\end{abstract}

\section{Introduction}\label{Sec:intro}

This work is part of a wider project that aims to reason about the correctness of code refactoring. To this end, a rigorous, formal definition is needed for the programming language under refactoring: in our case, Erlang. In earlier work, we developed a relational big-step semantics for sequential Core Erlang, including exceptions and side effects. This semantics is used in general proofs of characteristic properties (e.g. determinism) as well as proofs of equivalence between pairs of pattern expressions. The latter are important from the refactoring point of view: pattern equivalences can be interpreted as simple, correct refactorings for Core Erlang~\cite{bereczky_core,bereczky_machine}. Building on these simple equivalences, we plan to prove compound code transformations correct.

Formalising Core Erlang in the big-step operational definitional approach was a somewhat \emph{ad hoc} decision, supported by the following facts: it is not as detailed as small-step definitions, offering shorter proofs, and, at the same time, unlike in denotational definitions, semantics and proofs of nondeterministic and divergent programs do not need special treatment in the proof assistant embedding. Nonetheless, relational big-step semantics comes with its drawbacks: in general, it is not directly executable, the proof of determinism is complex, and we cannot use this style of semantics to argue about concurrency. After working with the relational big-step semantics formalisation for a while, the shortcomings of this approach became apparent, and we decided to investigate whether other semantics definition styles would be more suitable.

It seems to be a simple choice: the purpose of defining the semantics should determine the applied definition approach. However, conflicting requirements can make the decision unclear. For instance, in related work, different approaches have been applied to reason about program transformations: Grigore \emph{et al.}~\cite{langindepeq} and Garrido \emph{et al.}~\cite{4026866} use (reduction style) small-step semantics, whilst Owens \emph{et al.}~\cite{owens_functional} use (functional) big-step semantics. Both of these are executable and can be used to argue about program equivalence, but show different characteristics in general. There are a number of ways to create a testable and usable formal semantics, as, for example, addressed in a related discussion by Blazy and Leroy~\cite{blazy_machanized}, but it is not obvious to tell which is the best option for our purposes. Moreover, this choice is not only about the different mathematical approaches, but also how easy it is to implement them in the Coq proof assistant.

In this paper we analyse and compare different methods of defining big-step style semantics for a small, Erlang-like programming language. We do this to answer the question of which method should be used when creating a semantic description of sequential Core Erlang, when the description should support equivalence proofs and be efficiently executable. In doing this we survey the following methods: (a) traditional relational big-step semantics~\cite{kahn_natural}, (b) pretty-big-step semantics~\cite{pretty_big_step}, and (c) functional big-step semantics~\cite{owens_functional}, which can be seen as equivalent to supplying a definitional interpreter~\cite{reynolds_definitional}. We also briefly discuss a coinductive approach to define big-step semantics~\cite{leroy2008coinductive}. When comparing the semantic approaches, we aim to answer the following questions:

\begin{enumerate}
\item Does the semantics definition scale in terms of the complexity of expression equivalence proofs? Since our primary purpose is to prove expression pattern equivalences, the semantics has to be especially supportive of constructing such proofs.
\item Is the semantics effectively executable, allowing for automatic evaluation of expressions? Is this automatic execution efficient, with a performance comparable to a reference implementation? Execution of the semantics definition is crucial when it comes to validation: testing the semantics against a reference implementation needs the semantics to be executed.
\item How complex are the proofs for the common properties such as determinism or progress? For instance, some semantics are inherently deterministic, because they are presented as a semantic function, while it is a lot more cumbersome to prove  this property in a relational semantics.
\end{enumerate}

\noindent
We note that the paper not only makes a survey of the abovementioned semantics definition styles, but implements a benchmark language in each of those, and makes the detailed comparison based on the case study. Namely, we make the following main contributions:
\begin{itemize}
\item Traditional big-step, pretty-big-step and functional big-step semantics definitions for a simple functional programming language resembling sequential Core Erlang, moreover, we prove the equivalence of these definitions too. 
\item Proofs of basic properties of each semantics and proofs for simple expression pattern equivalences (local refactorings) in each definition style.
\item A systematic comparison of the approaches with respect to execution and proof complexity.
\end{itemize}

We will often quote Coq code to highlight the fact that all these concepts have been formalized in Coq~\cite{semantics_comp}. Inductive constructors in the relational semantics are described as inference rules.

The rest of the paper is structured as follows. In Section \ref{Sec:background} we describe the syntax, and necessary abstractions for our benchmark language. In Section \ref{Sec:relational} we discuss the traditional big-step and the pretty-big-step semantics, and in Section \ref{Sec:functional} we cover the functional approaches, in particular the functional big-step semantics. Section \ref{Sec:discussion} evaluates the presented approaches, and also briefly summarises coinductive big-step semantics~\cite{leroy2008coinductive}. Finally, Section \ref{Sec:conclusion} concludes and discusses future work. 

\section{The Benchmark Language}\label{Sec:background}

Throughout the paper, we define formal semantics for a simple but representative, functional programming language, which resembles Erlang; in fact, our case study language is a proper subset of Core Erlang. In this section, we introduce the syntax and a semantic domain for the language, based on which the later sections will define big-step operational semantics of different styles in order to make a systematic comparison between them.

\subsection{Syntax}

The case study language includes abstractions known from the functional paradigm (such as single assignment variables, \emph{let}-binding, lambda abstraction and function application), but we also incorporate impure expressions (such as I/O calls and exception handling). Furthermore, the language supports recursive function definitions (\emph{letrec}-binding), but only one name can be bound by each expression. Figure~\ref{Fig:syntax} defines the syntax of the language precisely, as an inductive type.

\begin{figure}[h!]
	
	\begin{minipage}{\textwidth}
		\coqdockw{Inductive} \coqdocvar{Expression} : \coqdockw{Type} :=\coqdoceol
		\coqdocnoindent
		\ensuremath{|} \coqdocvar{ELit} (\coqdocvar{l} : \coqdocvar{Literal}\footnote{Literals are either atoms or integers.})\coqdoceol
		\coqdocnoindent
		\ensuremath{|} \coqdocvar{EVar}     (\coqdocvar{v} : \coqdocvar{Var})\coqdoceol
		\coqdocnoindent
		\ensuremath{|} \coqdocvar{EFunId}  (\coqdocvar{f} : \coqdocvar{FunctionIdentifier})\coqdoceol
		\coqdocnoindent
		\ensuremath{|} \coqdocvar{EFun}     (\coqdocvar{vl} : \coqdocvar{list} \coqdocvar{Var}) (\coqdocvar{e} : \coqdocvar{Expression})\coqdoceol
		\coqdocnoindent
		\ensuremath{|} \coqdocvar{ECall}  (\coqdocvar{f} : \coqdocvar{string})     (\coqdocvar{params} : \coqdocvar{list} \coqdocvar{Expression}) \coqdoceol
		\coqdocnoindent
		\ensuremath{|} \coqdocvar{EApp} (\coqdocvar{exp} : \coqdocvar{Expression})     (\coqdocvar{params} : \coqdocvar{list} \coqdocvar{Expression})\coqdoceol
		\coqdocnoindent
		\ensuremath{|} \coqdocvar{ELet}   (\coqdocvar{v} : \coqdocvar{Var}) (\coqdocvar{e} \coqdocvar{b} : \coqdocvar{Expression})\coqdoceol
		\coqdocnoindent
		\ensuremath{|} \coqdocvar{ELetRec} (\coqdocvar{fid} : \coqdocvar{FunctionIdentifier}) (\coqdocvar{params} : \coqdocvar{list} \coqdocvar{Var}) (\coqdocvar{b} \coqdocvar{e} : \coqdocvar{Expression}) \coqdoceol
		\coqdocnoindent
		\ensuremath{|} \coqdocvar{ETry}   (\coqdocvar{e}$_1$  :  \coqdocvar{Expression}) (\coqdocvar{v} : \coqdocvar{Var}) (\coqdocvar{e}$_2$ : \coqdocvar{Expression}) (\coqdocvar{vl} : \coqdocvar{list} \coqdocvar{Var}) (\coqdocvar{e}$_3$ : \coqdocvar{Expression}).\coqdoceol
	\end{minipage}
	
	\caption{The syntax of our case study language (subset of Core Erlang)}
	\label{Fig:syntax}
\end{figure}

\subsection{Semantic Domain}

This language has expressions of three types: atoms, integers and functions. Therefore, values of expressions can only be literal values and closures (see Figure \ref{fig:value_syntax}). Closures are the normal forms of functions, and store the function's parameter list, body expression and an evaluation environment in which the body should be evaluated; moreover, the collection of recursive functions defined simultaneously\footnote{The presented approach is based on our previous work and is fairly general: it can handle multiple simultaneous function definitions, not only one; see \cite{bereczky_machine} for more details.}.

\begin{figure}[h]
    \begin{minipage}{\textwidth}
        \coqdockw{Inductive} \coqdocvar{Value} : \coqdockw{Type} :=\coqdoceol
        \coqdocnoindent
        \ensuremath{|} \coqdocvar{VLit} (\coqdocvar{l} : \coqdocvar{Literal})\coqdoceol
        \coqdocnoindent
        \ensuremath{|} \coqdocvar{VClos} (\coqdocvar{ref} : \coqdocvar{Environment}) (\coqdocvar{ext} : \coqdocvar{list} (\coqdocvar{FunctionIdentifier} \ensuremath{\times} \coqdocvar{FunctionExpression}))\coqdoceol
        \coqdocnoindent\hspace{33pt} (\coqdocvar{vl} : \coqdocvar{list} \coqdocvar{Var}) (\coqdocvar{e} : \coqdocvar{Expression}).\\ \coqdoceol
        \coqdockw{Definition} \coqdocvar{Exception} := \coqdocvar{ExceptionClass} $\times$ \coqdocvar{Value} $\times$ \coqdocvar{Value}.
    \end{minipage}
    \caption{Semantic domain}
    \label{fig:value_syntax}
\end{figure}

Exceptions may also be the results of expression evaluations. In our formalisation, exceptions are represented as triples: exception class (\verb|error|, \verb|throw| or \verb|exit|) and two values describing the exception reason. In our case studies, we will use two often seen exceptions known from Erlang: \coqdocvar{badarity} happens when an application evaluation fails due to the faulty parameter number, and \coqdocvar{badfun} is encountered when the main expression of the application evaluates to a value that is not a function closure.

Finally, we define the semantic domain as the union of values and exception descriptions: $\coqdocvar{Value} + \coqdocvar{Exception}$. In the formalisation, we use Coq's built-in union type with the standard \coqdocvar{inl} and \coqdocvar{inr} constructors to make elements of the semantic domain.

\subsection{Environment}

In order to share as much as possible in the different semantics definitions, not only we fix the semantic domain, but we define a common type for the evaluation environment. Basically, this is a collection of variable names (and function identifiers) mapped to values. There are several helper functions to manage this environment, namely:

\begin{itemize}
	\item \coqdocvar{get\_value}: Returns the value associated with a given name. If the name is unbound, it yields an exception.
	\item \coqdocvar{insert\_value}: Inserts a binding into the environment.
	\item \coqdocvar{append\_vars\_to\_env}: Inserts several variable bindings into the environment.
	\item \coqdocvar{append\_funs\_to\_env}: Inserts function identifier-closure bindings into the environment.
\end{itemize}

We remind the reader that the case study language allows for calling some built-in I/O functions, thus the semantics will need to address the meaning of these side-effects. For this, we define a type (\coqdocvar{SideEffectList}), the values of which log simple input-output effects produced by the evaluation of specific \coqdocvar{ECall} expressions. While evaluating \coqdocvar{ECall} expressions, we use the auxiliary \coqdocvar{eval} function, which returns a value or exception and a side effect trace --- only this operation can extend the side effect trace. In our previous work~\cite{bereczky_machine} we applied a slightly different method using standard list append operations in every derivation rule; however, this former approach had to be refined in order to support automatic evaluation of expressions.

\subsection{Evaluation Criteria}

As mentioned already in the introduction, we will compare the different approaches of defining big-step semantics based on the following criteria:

\begin{itemize}
	\item How complex is proving the properties of the semantics. We will use the determinism property to investigate this.
	\item Is the approach executable? Is the semantics efficiently executable?
	\item How complex is proving expression evaluation formally. We will use two smaller expressions to investigate this:
\begin{lstlisting}[caption={Expression evaluation examples},label={lst:examples},captionpos=b]
let X = fun(Y, Z) -> Y in 
  apply X('a', 'b')

let X = 4 in 
  let Y = 5 in 
    apply (fun(X, Y) -> X + Y) (X, Y) 
\end{lstlisting}
	\item How complex is proving expression equivalence? We will use one unconditional and one conditional\footnote{In the second example, the side effects produced by $e_1$ and $e_2$ are swapped during the evaluation of these expressions.} equivalence to investigate this:
\begin{lstlisting}[caption={Expression equivalence examples},label={lst:equivs},captionpos=b]
$e$ $\Leftrightarrow$ let $X$ = fun() -> $e$ in apply $X$()

let $A$ = $e_1$ in             let $B$ = $e_2$ in 
  let $B$ = $e_2$ in       $\Leftrightarrow$      let $A$ = $e_1$ in
    $A$ + $B$                    $A$ + $B$
\end{lstlisting}
\end{itemize}

\section{Relational Big-step Semantics}\label{Sec:relational}

A traditional big-step operational semantics is a relation between the evaluable expression and its value, or more generally, between initial and final configurations, where the configurations may include the evaluation environment or the side-effects of the evaluation. Note that in big-step style, the intermediate stages of the evaluation are not visible from the relation~\cite{nipkow2014concrete}. The idea of this style of semantics is originated from Kahn~\cite{kahn_natural}. In Coq, such a relation can be formalised with an inductive type, where the data constructors represent the derivation rules (or judgements).

\subsection{Traditional Relational Big-step Semantics}\label{Sec:inductive}

Traditional inductive big-step semantics are used in many projects, to mention but a few: deriving such a semantics from a small-step definition~\cite{ciob_smallbig}, call-by-need semantics of let and letrec calculus ($\lambda_{let}$, $\lambda_{letrec}$)~\cite{nakata_hasegawa_2009}, or the trace-based operational semantics for \emph{While}~\cite{nakata_trace-based} (this one is defined coinductively), as well as our project defining Core Erlang~\cite{bereczky_core,bereczky_machine,coreerlang}.

For the investigation of the different big-step definition styles, we reuse our Core Erlang formalisation mentioned above, but discard parts of it since the case study language used in the comparison is a subset of it. The big-step semantics will be denoted by $\bosse{\Gamma}{\coqdocvar{exp}}{\coqdocvar{eff}_1}{\coqdocvar{res}}{\coqdocvar{eff}_2}$ where $\Gamma$ is the evaluation environment, \coqdocvar{exp} is the evaluable expression, $\coqdocvar{eff}_1$ and $\coqdocvar{eff}_2$ are the initial and final side effect traces and \coqdocvar{res} is the result which is either a value or an exception. Before describing the semantics, we introduce some predicates and notations for readability about evaluating a list of expressions (we use $\lvert l \rvert$ to denote the length of list $l$,  $\coqdocvar{S i}$ denotes the successor of \coqdocvar{i} and $l[i]$ denotes the $i$th element of $l$). The function $\coqdocvar{nth\_def}\ \coqdocvar{l}\ \coqdocvar{default}\ \coqdocvar{i}$ works the same way as $\coqdocvar{l}[\coqdocvar{i} - 1]$ if $\coqdocvar{i} > 0$, but for $\coqdocvar{i} = 0$ it returns the $\coqdocvar{default}$ value. We also use Coq's standard \coqdocvar{last} function~\cite{coqdocref}.
\begin{flalign*}
&\textit{eval\_all}\ \Gamma\ \coqdocvar{es}\ \coqdocvar{vs}\ \coqdocvar{eff}\ \coqdocvar{eff}_1 :=
(\lvert \coqdocvar{es} \rvert = \lvert \coqdocvar{vs} \rvert) \Rightarrow (\lvert \coqdocvar{es} \rvert = \lvert \coqdocvar{eff} \rvert) \Rightarrow
(\forall \coqdocvar{j} < \lvert \coqdocvar{es} \rvert,\\
&\qquad\bosse{\Gamma}{\coqdocvar{es}[j]}{\coqdocvar{nth\_def}\ \coqdocvar{eff}\ \coqdocvar{eff}_1\ \coqdocvar{j}}{\inl{\coqdocvar{vs}[j]}}{\coqdocvar{nth\_def}\ \coqdocvar{eff}\ \coqdocvar{eff}_1\ (\coqdocvar{S j})})\\
&\textit{eval\_prefix}\ \Gamma\ \coqdocvar{es}\ \coqdocvar{vs}\ \coqdocvar{i}\ \coqdocvar{eff}\ \coqdocvar{eff}_1 := (i < \lvert \coqdocvar{es} \rvert) \Rightarrow (\lvert \coqdocvar{vs} \rvert = \coqdocvar{i}) \Rightarrow (\lvert \coqdocvar{eff} \rvert = \coqdocvar{i}) \Rightarrow\\
&\qquad(\forall \coqdocvar{j} < \coqdocvar{i}, \bosse{\Gamma}{\coqdocvar{es}[j]}{\coqdocvar{nth\_def}\ \coqdocvar{eff}\ \coqdocvar{eff}_1\ \coqdocvar{j}}{\inl{\coqdocvar{vs}[j]}}{\coqdocvar{nth\_def}\ \coqdocvar{eff}\ \coqdocvar{eff}_1\ (\coqdocvar{S j})}).
\end{flalign*}

The \textit{eval\_all} states that an expression list \coqdocvar{es} evaluates to a value list \coqdocvar{vs} (note that this formula also expresses that the evaluation of the expressions does not produce any exceptions). The evaluation for the $j$th step starts with the side effect log $\coqdocvar{eff}[j - 1]$ (or the default initial log $\coqdocvar{eff}_1$) and the result log is $\coqdocvar{eff}[j]$. The \textit{eval\_prefix} describes the same behaviour, but only for the first $i$ elements. Now we can describe the big-step semantics for our case study language (Figure \ref{Fig:traditional_big-step} shows the evaluation of expressions without exceptions, and Figure \ref{Fig:semantics_exception} explains the exceptional semantics).

\begin{figure*}[ht!]
	
	\centering
	In the following figures, the result \coqdocvar{res} could be either a value or an exception, so its type is $\coqdocvar{Value} + \coqdocvar{Exception}$.
	
	\begin{equation}
	\begin{prooftree}
	\infer0{\bosse{\Gamma}{\coqdocvar{ELit}\ \coqdocvar{l}}{\coqdocvar{eff}_1}{\inl{(\coqdocvar{VLit}\ \coqdocvar{l})}}{\coqdocvar{eff}_1}}
	\end{prooftree}
	\label{OS:lit}
	\tag{\textsc{Lit}}
	\end{equation}
	
	\begin{equation}
	\begin{prooftree}
	\hypo{\coqdocvar{res} = \coqdocvar{get\_value}\  \Gamma\ (\coqdocvar{inl}\ \coqdocvar{s})}
	\infer1{\bosse{\Gamma}{\coqdocvar{EVar}\ \coqdocvar{s}}{\coqdocvar{eff}_1}{\coqdocvar{res}}{\coqdocvar{eff}_1}}
	\end{prooftree}
	\label{OS:var}
	\tag{\textsc{Var}}
	\end{equation}
	
	\begin{equation}
	\begin{prooftree}
	\infer0{\bosse{\Gamma}{\coqdocvar{EFun}\ \coqdocvar{vl}\ \coqdocvar{e}}{\coqdocvar{eff}_1}{\inl{(\coqdocvar{VClos}\ \Gamma\ \langle \rangle\ \coqdocvar{vl}\ \coqdocvar{e})}}{\coqdocvar{eff}_1}}
	\end{prooftree}
	\label{OS:fun}
	\tag{\textsc{Fun}}
	\end{equation}

	\begin{equation}
	\begin{prooftree}
	\hypo{\coqdocvar{res} = \coqdocvar{get\_value}\ \Gamma\ (\coqdocvar{inr}\ \coqdocvar{fid})}
	\infer1{\bosse{\Gamma}{\coqdocvar{EFunId}\ \coqdocvar{fid}}{\coqdocvar{eff}_1}{\coqdocvar{res}}{\coqdocvar{eff}_1}}
	\end{prooftree}
	\label{OS:funid}
	\tag{\textsc{FunId}}
	\end{equation}
	
	\begin{equation}
	\begin{prooftree}
	\hypo{\textit{eval\_all}\ \Gamma\ \coqdocvar{params}\ \coqdocvar{vals}\ \coqdocvar{eff}_1\ \coqdocvar{eff}}
	\hypo{\coqdocvar{eval}\ \coqdocvar{fname}\ \coqdocvar{vals}\ (\coqdocvar{last}\ \coqdocvar{eff}\ \coqdocvar{eff}_1) = (\coqdocvar{res}, \coqdocvar{eff}_2)}
	\infer2{\bosse{\Gamma}{\coqdocvar{ECall}\ \coqdocvar{fname}\ \coqdocvar{params}}{\coqdocvar{eff}_1}{\coqdocvar{res}}{\coqdocvar{eff}_2}}
	\end{prooftree}
	\label{OS:call}
	\tag{\textsc{Call}}
	\end{equation}
	
	\begin{equation}
	\begin{prooftree}
	\hypo{\bosse{\Gamma}{\coqdocvar{exp}}{\coqdocvar{eff}_1}{\inl{(\coqdocvar{VClos}\ \coqdocvar{ref}\ \coqdocvar{ext}\ \coqdocvar{var\_list}\ \coqdocvar{body})}}{\coqdocvar{eff}_2}}
	\hypo{\lvert \coqdocvar{var\_list} \rvert = \lvert \coqdocvar{vals} \rvert}
	\infer[no rule]1{\textit{eval\_all}\ \Gamma\ \coqdocvar{params}\ \coqdocvar{vals}\ \coqdocvar{eff}_2\ \coqdocvar{eff}}
	\infer[no rule]2{\bosse{\coqdocvar{append\_vars\_to\_env}\ \coqdocvar{var\_list}\ \coqdocvar{vals}\ (\coqdocvar{get\_env}\ \coqdocvar{ref}\ \coqdocvar{ext})}{\coqdocvar{body}}{\coqdocvar{last}\ \coqdocvar{eff}\ \coqdocvar{eff}_2}{\coqdocvar{res}}{\coqdocvar{eff}_3}}
	\infer1{\bosse{\Gamma}{\coqdocvar{EApp}\ \coqdocvar{exp}\ \coqdocvar{params}}{\coqdocvar{eff}_1}{\coqdocvar{res}}{\coqdocvar{eff}_3}}
	\end{prooftree}
	\label{OS:apply}
	\tag{\textsc{App}}
	\end{equation}
	
	\begin{equation}
	\begin{prooftree}
	\hypo{\bosse{\Gamma}{\coqdocvar{e}}{\coqdocvar{eff}_1}{\inlcv{val}}{\coqdocvar{eff}_2}}
	\hypo{\bosse{\coqdocvar{insert\_value}\ \Gamma\ \coqdocvar{v}\ \coqdocvar{val}}{\coqdocvar{b}}{\coqdocvar{eff}_2}{\coqdocvar{res}}{\coqdocvar{eff}_3}} 
	\infer2{\bosse{\Gamma}{\coqdocvar{ELet}\ \coqdocvar{v}\ \coqdocvar{e}\ \coqdocvar{b}}{\coqdocvar{eff}_1}{\coqdocvar{res}}{\coqdocvar{eff}_3}}
	\end{prooftree}
	\label{OS:let}
	\tag{\textsc{Let}}
	\end{equation}
	
	\begin{equation}
	\begin{prooftree}
	\hypo{\bosse{\coqdocvar{append\_funs\_to\_env}\ [\coqdocvar{fid}]\ [\coqdocvar{params}]\ [\coqdocvar{b}]\ \Gamma}{\coqdocvar{e}}{\coqdocvar{eff}_1}{\coqdocvar{res}}{\coqdocvar{eff}_2}} 
	\infer1{\bosse{\Gamma}{\coqdocvar{ELetRec}\ \coqdocvar{fid}\ \coqdocvar{params}\ \coqdocvar{b}\ \coqdocvar{e}}{\coqdocvar{eff}_1}{\coqdocvar{res}}{\coqdocvar{eff}_2}}
	\end{prooftree}
	\label{OS:letrec}
	\tag{\textsc{LetRec}}
	\end{equation}
	
	\caption{The core traditional big-step definition of our case study language}
	\label{Fig:traditional_big-step}
\end{figure*}

\begin{figure}[ht!]
	\centering
	\begin{equation}
	\begin{prooftree}
	\hypo{\bosse{\Gamma}{\coqdocvar{e}_1}{\coqdocvar{eff}_1}{\inlcv{val'}}{\coqdocvar{eff}_2}}
	\hypo{\bosse{\coqdocvar{insert\_value}\ \Gamma\ \coqdocvar{v}\ \coqdocvar{val'}}{\coqdocvar{e}_2}{\coqdocvar{eff}_2}{\coqdocvar{res}}{\coqdocvar{eff}_3}}
	\infer2{\bosse{\Gamma}{\coqdocvar{ETry}\ \coqdocvar{e}_1\ \coqdocvar{v}\ \coqdocvar{e}_2\ \coqdocvar{vl}\ \coqdocvar{e}_3}{\coqdocvar{eff}_1}{\coqdocvar{res}}{\coqdocvar{eff}_3}}
	\label{OS:try}
	\tag{\textsc{Try}}
	\end{prooftree}
	\end{equation}

	\begin{equation}
	\begin{prooftree}
	\hypo{\bosse{\Gamma}{\coqdocvar{e}_1}{\coqdocvar{eff}_1}{\inr{(\coqdocvar{ex}_1, \coqdocvar{ex}_2, \coqdocvar{ex}_3)}}{\coqdocvar{eff}_2}}
	\infer[no rule]1{\bosse{\coqdocvar{append\_try\_vars\_to\_env}\ \coqdocvar{vl}\ [\coqdocvar{exclass\_to\_value}\ \coqdocvar{ex}_1; \coqdocvar{ex}_2; \coqdocvar{ex}_3]\ \Gamma}{\coqdocvar{e}_3}{\coqdocvar{eff}_2}{\coqdocvar{res}}{\coqdocvar{eff}_3}}
	\infer1{\bosse{\Gamma}{\coqdocvar{ETry}\ \coqdocvar{e}_1\ \coqdocvar{v}\ \coqdocvar{e}_2\ \coqdocvar{vl}\ \coqdocvar{e}_3}{\coqdocvar{eff}_1}{\coqdocvar{res}}{\coqdocvar{eff}_3}}
	\label{OS:catch}
	\tag{\textsc{Catch}}
	\end{prooftree}
	\end{equation}
	
	For the next rule, let us consider $\textit{nonclosure}\ \coqdocvar{v} := \forall\ \Gamma', \coqdocvar{ext}, \coqdocvar{var\_list}, \coqdocvar{body}: \coqdocvar{v} \neq \coqdocvar{VClos}\ \Gamma'\ \coqdocvar{ext}\ \coqdocvar{var\_list}\ \coqdocvar{body}$.
	
	\begin{equation}
	\begin{prooftree}
	\hypo{\bosse{\Gamma}{\coqdocvar{exp}}{\coqdocvar{eff}_1}{\inlcv{v}}{\coqdocvar{eff}_2}}
	\infer[no rule]1{\textit{eval\_all}\ \Gamma\ \coqdocvar{params}\ \coqdocvar{vals}\ \coqdocvar{eff}_2\ \coqdocvar{eff}}
	\hypo{\textit{nonclosure}\ \coqdocvar{v}}
	\hypo{\coqdocvar{eff}_3 = \coqdocvar{last}\ \coqdocvar{eff}\ \coqdocvar{eff}_2}
	\infer3{\bosse{\Gamma}{\coqdocvar{EApp}\ \coqdocvar{exp}\ \coqdocvar{params}}{\coqdocvar{eff}_1}{\inr{(\coqdocvar{badfun}\ \coqdocvar{v})}}{\coqdocvar{eff}_3}}
	\end{prooftree}
	\label{OS:apply_ex_badfun}
	\tag{\textsc{AppExc$_1$}}
	\end{equation}
	
	In the following rule, we denote $\coqdocvar{VClos}\ \coqdocvar{ref}\ \coqdocvar{ext}\ \coqdocvar{var\_list}\ \coqdocvar{body}$ with $v$.
	
	\begin{equation}
	\begin{prooftree}
	\hypo{\bosse{\Gamma}{\coqdocvar{exp}}{\coqdocvar{eff}_1}{\inl{v}}{\coqdocvar{eff}_2}}
	\infer[no rule]1{\textit{eval\_all}\ \Gamma\ \coqdocvar{params}\ \coqdocvar{vals}\ \coqdocvar{eff}_2\ \coqdocvar{eff}}
	\hypo{\lvert \coqdocvar{var\_list} \rvert \neq \lvert \coqdocvar{vals} \rvert}
	\infer[no rule]1{\coqdocvar{eff}_3 = \coqdocvar{last}\ \coqdocvar{eff}\ \coqdocvar{eff}_2}
	\infer2{\bosse{\Gamma}{\coqdocvar{EApp}\ \coqdocvar{exp}\ \coqdocvar{params}}{\coqdocvar{eff}_1}{\inr{(\coqdocvar{badarity}\ \coqdocvar{v})}}{\coqdocvar{eff}_3}}
	\end{prooftree}
	\label{OS:apply_ex_badarity}
	\tag{\textsc{AppExc$_2$}}
	\end{equation}
	
	\begin{equation}
	\begin{prooftree}
	\hypo{\textit{eval\_prefix}\ \Gamma\ \coqdocvar{params}\ \coqdocvar{vals}\ \coqdocvar{i}\ \coqdocvar{eff}_1\ \coqdocvar{eff}}
	\hypo{\bosse{\Gamma}{\coqdocvar{params}[i]}{\coqdocvar{last}\ \coqdocvar{eff}\ \coqdocvar{eff}_1}{\inrcv{ex}}{\coqdocvar{eff}_2}}
	\infer2{\bosse{\Gamma}{\coqdocvar{ECall}\ \coqdocvar{fname}\ \coqdocvar{params}}{\coqdocvar{eff}_1}{\inrcv{ex}}{\coqdocvar{eff}_2}}
	\end{prooftree}
	\label{OS:call_ex}
	\tag{\textsc{CallExc}}
	\end{equation}

	\begin{equation}
	\begin{prooftree}
	\hypo{\bosse{\Gamma}{\coqdocvar{e}}{\coqdocvar{eff}_1}{\inrcv{ex}}{\coqdocvar{eff}_2}}
	\infer1{\bosse{\Gamma}{\coqdocvar{ELet}\ \coqdocvar{v}\ \coqdocvar{e}\ \coqdocvar{b}}{\coqdocvar{eff}_1}{\inrcv{ex}}{\coqdocvar{eff}_2}}
	\end{prooftree}
	\label{OS:let_ex}
	\tag{\textsc{LetExc}}
	\end{equation}
	
	\begin{equation}
	\begin{prooftree}
	\hypo{\bosse{\Gamma}{\coqdocvar{exp}}{\coqdocvar{eff}_1}{\inrcv{ex}}{\coqdocvar{eff}_2}}
	\infer1{\bosse{\Gamma}{\coqdocvar{EApp}\ \coqdocvar{exp}\ \coqdocvar{params}}{\coqdocvar{eff}_1}{\inrcv{ex}}{\coqdocvar{eff}_2}}
	\end{prooftree}
	\label{OS:apply_ex_exp}
	\tag{\textsc{AppExc$_3$}}
	\end{equation}
	
	\begin{equation}
	\begin{prooftree}
	\hypo{\bosse{\Gamma}{\coqdocvar{exp}}{\coqdocvar{eff}_1}{\inlcv{v}}{\coqdocvar{eff}_2}}
	\hypo{\textit{eval\_prefix}\ \Gamma\ \coqdocvar{params}\ \coqdocvar{vals}\ \coqdocvar{i}\ \coqdocvar{eff}_2\ \coqdocvar{eff}}
	\infer[no rule]1{\bosse{\Gamma}{\coqdocvar{params}[\coqdocvar{i}]}{\coqdocvar{last}\ \coqdocvar{eff}\ \coqdocvar{eff}_2}{\inrcv{ex}}{\coqdocvar{eff}_3}}
	\infer2{\bosse{\Gamma}{\coqdocvar{EApp}\ \coqdocvar{exp}\ \coqdocvar{params}}{\coqdocvar{eff}_1}{\inrcv{ex}}{\coqdocvar{eff}_3}}
	\end{prooftree}
	\label{OS:apply_ex_param}
	\tag{\textsc{AppExc$_4$}}
	\end{equation}
	
	\caption{The traditional big-step operational semantics of exception creation and propagation}
	\label{Fig:semantics_exception}
\end{figure}

\subsubsection{Making it Executable}

The traditional, relational big-step semantics introduced in the previous section is not inherently executable or computable: for a given pair of starting and final configurations, it needs to be proven that they are in operational semantics relation. In Coq, such a proof can be given in terms of proof primitives, or one can write a program in the tactic language to construct the proof. Automatic execution of relational semantics can be done with the latter. We could also see the evaluation tactics as a machinery that can turn the relational semantics into functional: a program in the tactic language can perform pattern matching, case distinction and even recursion, and can ultimately compute the results of the relation.

\paragraph{Limitations of Coq tactics.}

Executing the relational semantics in Coq involves technical considerations: one needs to make sure that the operational semantics derivation rules do not contain auxiliary function calls in their consequences. Otherwise, the Coq tactic language cannot do simple pattern matching on the proof goals and prevents syntax-directed evaluation. In our semantics, we needed to apply minor changes in the derivation rule of variables and at uses of the \emph{append} operation on side effect logs in our Core Erlang semantics~\cite{bereczky_machine}. The issue has been solved by refactoring: we replaced the auxiliary function applications with fresh variables and added extra premises stating equality between the variables and the corresponding function applications.

On the other hand, in case of the side effects (and the mentioned append operations) to avoid the introduction of unreasonable numbers of new variables, we changed the use of these traces. Note that currently only \coqdocvar{ECall} expressions can cause new side effects, the other rules just have to propagate the logs. Instead of handling only the additional side effects of an expression evaluation step, we rather consider using always the whole initial and final side effect traces (i.e. not only the difference like in~\cite{bereczky_machine}). This way we could dispose of the append operations in the consequences of the derivation rules.

\paragraph{Evaluation tactic.}

We use Coq's tactic sublanguage called Ltac~\cite{coqltac} to automate proof construction. In our case, the evaluation of the semantics of the case study language without exceptions is syntax-directed, i.e. a tactic can be designed to evaluate any expression in any context based on pattern-matching on the expression to be evaluated (e.g. \coqdocvar{ECall} expression can be evaluated with \ref{OS:call}). On the other hand, after introducing exceptions, several derivation rules are applicable for evaluating a particular expression (e.g. there are two rules for \coqdocvar{ECall}, five rules for function applications, etc.). We extended the evaluation tactic to try applying the applicable rules one after the other. This can be seen as a backtracking proof-search for a successful evaluation path.

As it turned out, such evaluation tactics in Coq are rather ineffective in terms of time and space. To speed up the execution, we can create some helper functions and prove lemmas about specific expressions (e.g. the evaluation of parameters which are just literals), so that the evaluation tactic can apply these lemmas before trying to evaluate an expression with the mentioned slow backtracking process. These lemmas can significantly speed up the evaluation of expressions which contain such specific sub-expressions; however, they only solve a small part of the problem.

\subsection{Pretty-Big-step Semantics}\label{Sec:pretty}

As seen before, the traditional definition contains several similar rules with the same premises. The idea of Charguéraud --- called pretty-big-step semantics~\cite{pretty_big_step} --- is focusing on eliminating this redundancy. Let us discuss his idea through our case study using the evaluation rules for applications. First of all, Charguéraud identified two sources of duplication:

\begin{itemize}
	\item The similar premises in the rules for exceptions, correct evaluation (and divergence).
	\item The duplication of the evaluation judgement both for values and exceptions. This is not present in our case study language, however, \ref{OS:apply} could be described in form of two rules: one for exception and one for the value final result.
\end{itemize}

In the following paragraphs we focus on the first problem. Instead of using duplicated conditions, Charguéraud suggests to use ``intermediate terms'' which contain the satisfied conditions implicitly. These can be seen also as terms, which remember the state of the evaluation, i.e. which sub-terms have already been evaluated (this resembles a small-step semantics in some aspects).

\paragraph{Applications with intermediate terms.}

Let us see how the idea applies to our semantics. First, we need to create the syntax for intermediate terms (see Figure \ref{Fig:pretty_syntax}). In our case, we need three additional constructors for applications: \coqdocvar{AApp1} corresponds to the function expression evaluation, \coqdocvar{AList} to the evaluation of the parameters, while \coqdocvar{AApp2} to the application exception creation and function body evaluation.

\begin{figure}[h]
	\coqdockw{Inductive} \coqdocvar{AuxExpression} :=\coqdoceol\coqdocnoindent
	\ensuremath{|} \coqdocvar{AApp1} (\coqdocvar{b} : \coqdocvar{Value} \ensuremath{+} \coqdocvar{Exception}) (\coqdocvar{params} : \coqdocvar{list} \coqdocvar{Expression})\coqdoceol\coqdocnoindent
	\ensuremath{|} \coqdocvar{AApp2} (\coqdocvar{v} : \coqdocvar{Value}) (\coqdocvar{b} : \coqdocvar{list} \coqdocvar{Value} \ensuremath{+} \coqdocvar{Exception})\coqdoceol
	$\cdots$\coqdoceol.\\
	
	\coqdockw{Inductive} \coqdocvar{AuxList} := \coqdocvar{AList} (\coqdocvar{rest} : \coqdocvar{list} \coqdocvar{Expression}) (\coqdocvar{b} : \coqdocvar{list} \coqdocvar{Value} \ensuremath{+} \coqdocvar{Exception}).
	\caption{The syntax of intermediate terms}
	\label{Fig:pretty_syntax}
\end{figure}

After having the intermediate terms defined, we can rewrite the semantics of applications (Figure \ref{Fig:pretty_semantics}). We decided not to include our side effect traces in the intermediate terms, because this way the effects can be handled just like before. First, we have to evaluate the function expression of the application (\ref{OS:pretty_app_1}). We create the intermediate term \coqdocvar{AApp1} with the result of this step. If this result was an exception, then the evaluation is finished with \ref{OS:pretty_app_1_exc}, otherwise, the parameters follow after using \ref{OS:pretty_app_1_fin}.

When there are parameters, we can take the first one and evaluate it with \ref{OS:pretty_list_step}. The result will be appended to the end of the value list in the constructor \coqdocvar{AList} if it is a value by the \coqdocvar{mk\_result} function; however, in case of an exception this attribute of \coqdocvar{AList} becomes the mentioned exception. We repeat this process until all parameter expressions are evaluated, or an exception occurs inside the \coqdocvar{AList}. In the latter case, \ref{OS:pretty_list_exc} finishes the evaluation, and the stored exception will be propagated. When there is no exception, we use \ref{OS:pretty_list_fin} to finish the parameter list evaluation. At this point, we can notice that this a general approach to evaluating a list of expressions, so it can be used for \coqdocvar{ECall} expressions too.

\begin{figure}[ht!]
	\centering
    
    In this figure, \coqdocvar{lres} is either a list of \coqdocvar{Value}s, or an exception, so its type is \coqdocvar{list} \coqdocvar{Value} + \coqdocvar{Exception}.
	
	\begin{equation}
	\begin{prooftree}
	\hypo{\pbosse{\Gamma}{\coqdocvar{exp}}{\coqdocvar{eff}_1}{\coqdocvar{res}_1}{\coqdocvar{eff}_2}}
	\hypo{\pbosse{\Gamma}{\coqdocvar{AApp1}\ \coqdocvar{res}_1\ \coqdocvar{params}}{\coqdocvar{eff}_2}{\coqdocvar{res}_2}{\coqdocvar{eff}_3}}
	\infer2{\pbosse{\Gamma}{\coqdocvar{EApp}\ \coqdocvar{exp}\ \coqdocvar{params}}{\coqdocvar{eff}_1}{\coqdocvar{res}_2}{\coqdocvar{eff}_3}}
	\end{prooftree}
	\label{OS:pretty_app_1}
	\tag{\textsc{App$_1^{\textit{pretty}}$}}
	\end{equation}
	
	\begin{equation}
	\begin{prooftree}
	\infer0{\pbosse{\Gamma}{\coqdocvar{AApp1}\ (\inrcv{ex})\ \coqdocvar{params}}{\coqdocvar{eff}_1}{\inrcv{ex}}{\coqdocvar{eff}_1}}
	\end{prooftree}
	\label{OS:pretty_app_1_exc}
	\tag{\textsc{ExcApp$_1^{\textit{pretty}}$}}
	\end{equation}
	
	\begin{equation}
	\begin{prooftree}
	\hypo{\pbosse{\Gamma}{\coqdocvar{AList}\ \coqdocvar{params}\ (\inl{[]})}{\coqdocvar{eff}_1}{\coqdocvar{lres}}{\coqdocvar{eff}_2}}
	\infer[no rule]1{\pbosse{\Gamma}{\coqdocvar{AApp2}\ \coqdocvar{v}\ \coqdocvar{lres}}{\coqdocvar{eff}_2}{\coqdocvar{res}}{\coqdocvar{eff}_3}}
	\infer1{\pbosse{\Gamma}{\coqdocvar{AApp1}\ (\inlcv{v})\ \coqdocvar{params}}{\coqdocvar{eff}_1}{\coqdocvar{res}}{\coqdocvar{eff}_3}}
	\end{prooftree}
	\label{OS:pretty_app_1_fin}
	\tag{\textsc{FinApp$_1^{\textit{pretty}}$}}
	\end{equation}
	
	\begin{equation}
	\begin{prooftree}
	\infer0{\pbosse{\Gamma}{\coqdocvar{AApp2}\ \coqdocvar{v}\ (\inrcv{ex})}{\coqdocvar{eff}_1}{\inrcv{ex}}{\coqdocvar{eff}_1}}
	\end{prooftree}
	\label{OS:pretty_app_2_exc}
	\tag{\textsc{ExcApp$_2^{\textit{pretty}}$}}
	\end{equation}
	
	\begin{equation}
	\begin{prooftree}
	\hypo{\lvert \coqdocvar{var\_list} \rvert = \lvert \coqdocvar{vals} \rvert}
	\infer[no rule]1{\pbosse{\coqdocvar{append\_vars\_to\_env}\ \coqdocvar{var\_list}\ \coqdocvar{vals}\ (\coqdocvar{get\_env}\ \coqdocvar{ref}\  \coqdocvar{ext})}{\coqdocvar{body}}{\coqdocvar{eff}_1}{\coqdocvar{res}}{\coqdocvar{eff}_2}}
	\infer1{\pbosse{\Gamma}{\coqdocvar{AApp2}\ (\coqdocvar{VClos}\ \coqdocvar{ref}\ \coqdocvar{ext}\ \coqdocvar{var\_list}\ \coqdocvar{body})\ (\inlcv{vals})}{\coqdocvar{eff}_1}{\coqdocvar{res}}{\coqdocvar{eff}_2}}
	\end{prooftree}
	\label{OS:pretty_app_2_fin}
	\tag{\textsc{FinApp$_2^{\textit{pretty}}$}}
	\end{equation}
	
	\begin{equation}
	\begin{prooftree}
	\hypo{\textit{nonclosure}\ v}
	\infer1{\pbosse{\Gamma}{\coqdocvar{AApp2}\ v\ (\inlcv{vals})}{\coqdocvar{eff}_1}{\inr{(\coqdocvar{badfun}\ \coqdocvar{v})}}{\coqdocvar{eff}_1}}
	\end{prooftree}
	\label{OS:pretty_app_exc_1}
	\tag{\textsc{ExcApp$_{2, \textit{badfun}}^{\textit{\textit{pretty}}}$}}
	\end{equation}
	
	In the following rule, we denote $\coqdocvar{VClos}\ \coqdocvar{ref}\ \coqdocvar{ext}\ \coqdocvar{var\_list}\ \coqdocvar{body}$ with $v$.
	
	\begin{equation}
	\begin{prooftree}
	\hypo{\lvert \coqdocvar{var\_list} \rvert \neq \lvert \coqdocvar{vals} \rvert}
	\infer1{\pbosse{\Gamma}{\coqdocvar{AApp2}\ v\ (\inlcv{vals})}{\coqdocvar{eff}_1}{\inr{(\coqdocvar{badarity}\ v)}}{\coqdocvar{eff}_1}}
	\end{prooftree}
	\label{OS:pretty_app_exc_2}
	\tag{\textsc{ExcApp$_{2, \textit{badarity}}^{\textit{pretty}}$}}
	\end{equation}
	
	\begin{equation}
	\begin{prooftree}
	\infer0{\pbosse{\Gamma}{\coqdocvar{AList}\ []\ (\inlcv{vals})}{\coqdocvar{eff}_1}{\inlcv{vals}}{\coqdocvar{eff}_1}}
	\end{prooftree}
	\label{OS:pretty_list_fin}
	\tag{\textsc{FinList$^{\textit{pretty}}$}}
	\end{equation}
	
	\begin{equation}
	\begin{prooftree}
	\infer0{\pbosse{\Gamma}{\coqdocvar{AList}\ \coqdocvar{rest}\ (\inrcv{ex})}{\coqdocvar{eff}_1}{\inrcv{ex}}{\coqdocvar{eff}_1}}
	\end{prooftree}
	\label{OS:pretty_list_exc}
	\tag{\textsc{ExcList$^{\textit{pretty}}$}}
	\end{equation}
	
	\begin{equation}
	\begin{prooftree}
	\hypo{\pbosse{\Gamma}{\coqdocvar{r}}{\coqdocvar{eff}_1}{\coqdocvar{res}}{\coqdocvar{eff}_2}}
	\hypo{\pbosse{\Gamma}{\coqdocvar{AList}\ \coqdocvar{rest}\ (\coqdocvar{mk\_result}\ \coqdocvar{res}\ \coqdocvar{vals})}{\coqdocvar{eff}_2}{\coqdocvar{lres}}{\coqdocvar{eff}_3}}
	\infer2{\pbosse{\Gamma}{\coqdocvar{AList}\ (\coqdocvar{r} ::\coqdocvar{rest})\ (\inlcv{vals})}{\coqdocvar{eff}_1}{\coqdocvar{lres}}{\coqdocvar{eff}_3}}
	\end{prooftree}
	\label{OS:pretty_list_step}
	\tag{\textsc{StepList$^{\textit{pretty}}$}}
	\end{equation}
	
	\caption{Pretty-big-step semantics for applications}
	\label{Fig:pretty_semantics}
\end{figure}

Finally, if there was an exception during parameter evaluation, instead of the parameter values, an exception is stored in \coqdocvar{AApp2}, and this can be propagated with \ref{OS:pretty_app_2_exc}. Otherwise, all parameters were correctly evaluated and \ref{OS:pretty_app_2_fin} can be applied, when the first saved value (the evaluated application function expression) is a closure, moreover, the number of formal parameters in this closure is the same as the actual parameters, which are also stored in a value list in \coqdocvar{AApp2}.  However, if the first stored value is not a closure, we use \ref{OS:pretty_app_exc_1} and a \coqdocvar{badfun} exception will be the result, otherwise, we can apply \ref{OS:pretty_app_exc_2} if the number of formal and actual parameters mismatch to create a \coqdocvar{badarity} exception.

\paragraph{Brief evaluation.}

Compared to the traditional semantics, in the pretty-big-step approach we see the increase in the number of inference rules, while the premise redundancy is eliminated and the number of premises drops to two at most. Obviously, pretty-big-step semantics cannot overcome all weaknesses of the big-step approach, but provides a good alternative in terms of readability and usability.

Transforming the big-step semantics to pretty-big-step style was a straightforward process, except the transformation of expression lists: if there were two or more derivation steps in the big-step premises, the intermediate results were turned into terms like \coqdocvar{AApp1}, and the rule was split. These steps could have been automated, however, in case of expression lists, the use of accumulation in \coqdocvar{AList} instead of the \textit{eval\_all} predicate was not as simple.

We should also note, that the pretty-big-step definition is a relational semantics just like the traditional one. This means, we need a tactic again to execute this semantics, however, unlike in the traditional case, here backtracking is not needed, because the evaluation is syntax-driven (with the exception of the last step of application evaluation: \ref{OS:pretty_app_2_fin}, \ref{OS:pretty_app_exc_1} and \ref{OS:pretty_app_exc_2}). Although, the use of this evaluation tactic is still not efficient enough.

There are interesting applications of the method in related research, such as deriving pretty-big-step style semantics from small-step semantics~\cite{poulsen_deriving} or certified abstract interpretation using this definition style~\cite{bodin_certified}.

\section{Functional Ways to Define Big-step Semantics}\label{Sec:functional}

We have summarised two ways to create a relational big-step semantics, however, both of them suffer mainly from the same problem: they cannot be executed efficiently. In this section, we discuss a functional approach, called functional big-step semantics~\cite{owens_functional} and its origin, the definitional interpreter~\cite{reynolds_definitional}.

\subsection{Functional Big-Step Semantics}\label{Sec:Functional_big_step}

The idea of functional big-step semantics was developed by Owens \emph{et al.}~\cite{owens_functional}. A semantics in this style is basically a recursive function. In order to assure its termination for arbitrary inputs (e.g. for diverging expressions too), there is also a ``clock'' variable which decreases in the steps of the execution. We note that the functional big-step semantics is essentially a definitional interpreter~\cite{reynolds_definitional} equipped with a clock, and it is defined in a ``higher-order logic rather than a programming language''~\cite{owens_functional}.

This approach is also used in research, for example in the FEther project~\cite{yang_feether} and the type soundness proof for System F by Amin and Rompf~\cite{nada_type} uses definitional interpreters, while a verified compiler backend for CakeML~\cite{owens_functional,Yong_cakeml} in based on functional big-step semantics. Now, let us see how can we create such a semantics for our case study language.

When we define a function (in Coq), it should explicitly implement behaviour for all inputs (i.e. the function is total). However, in practice, there can be programs or expressions with undefined or unspecified behaviour (naturally, that the relational semantics are partial too). Moreover, because of the ``clock'' variable which ensures the termination of the function, an expression evaluation could terminate before finding the right result for small ``clocks''. This means, we can have three different results: \emph{correct termination}, \emph{failure}, and \emph{timeout} (see Figure \ref{Fig:functional_values}). In our case study language, there is no undefined behaviour, so we never get \emph{failure} as result, however, we do not omit this from the result definition, because if we extend this definition e.g. with Core Erlang-like \verb|case| expressions, then the guards of these expressions cannot produce observable side effects~\cite{carlsson2000core}, so the semantics of \verb|case| expressions with such guards would be \emph{failure}.

\begin{figure}[t]
    \coqdockw{Inductive} \coqdocvar{ResultType} : \coqdockw{Type} :=\coqdoceol
    \coqdocnoindent
    \ensuremath{|} \coqdocvar{Result} (\coqdocvar{res} : \coqdocvar{Value} + \coqdocvar{Exception}) (\coqdocvar{eff} : \coqdocvar{SideEffectList})\coqdoceol
    \coqdocnoindent
    \ensuremath{|} \coqdocvar{Timeout}\coqdoceol
    \coqdocnoindent
    \ensuremath{|} \coqdocvar{Failure}.
    \\

    \coqdockw{Inductive} \coqdocvar{ResultListType} : \coqdockw{Type} :=\coqdoceol
    \coqdocnoindent
    \ensuremath{|} \coqdocvar{LResult} (\coqdocvar{res} : \coqdocvar{list} \coqdocvar{Value} + \coqdocvar{Exception}) (\coqdocvar{eff} : \coqdocvar{SideEffectList})\coqdoceol
    \coqdocnoindent
    \ensuremath{|} \coqdocvar{LTimeout}\coqdoceol
    \coqdocnoindent
    \ensuremath{|} \coqdocvar{LFailure}.
    \caption{The possible results of the functional big-step semantics}
    \label{Fig:functional_values}
\end{figure}

\begin{figure}[t]
    \coqdockw{Fixpoint} \coqdocvar{eval\_elems} (\coqdocvar{f} : \coqdocvar{Environment} \ensuremath{\rightarrow} \coqdocvar{Expression} \ensuremath{\rightarrow} \coqdocvar{SideEffectList} \ensuremath{\rightarrow} \coqdocvar{ResultType})\coqdoceol
    \coqdocindent{0.50em}($\Gamma$ : \coqdocvar{Environment}) (\coqdocvar{exps} : \coqdocvar{list} \coqdocvar{Expression}) (\coqdocvar{eff} : \coqdocvar{SideEffectList}) : \coqdocvar{ResultListType} := \coqdoceol
    \coqdocnoindent
    \coqdockw{match} \coqdocvar{exps} \coqdockw{with}\coqdoceol
    \coqdocnoindent
    \ensuremath{|} []    \ensuremath{\Rightarrow} \coqdocvar{LResult} (\coqdocvar{inl} []) \coqdocvar{eff}\coqdoceol
    \coqdocnoindent
    \ensuremath{|} \coqdocvar{x}::\coqdocvar{xs} \ensuremath{\Rightarrow} \coqdoceol
    \coqdocindent{1.00em}
    \coqdockw{match} \coqdocvar{f} $\Gamma$  \coqdocvar{x} \coqdocvar{eff} \coqdockw{with}\coqdoceol
    \coqdocindent{1.00em}
    \ensuremath{|} \coqdocvar{Result} (\coqdocvar{inl} \coqdocvar{v}) \coqdocvar{eff'} \ensuremath{\Rightarrow} 
    \coqdockw{let} \coqdocvar{res} := \coqdocvar{eval\_elems} \coqdocvar{f} $\Gamma$ \coqdocvar{xs} \coqdocvar{eff'} \coqdoctac{in}\coqdoceol
    \coqdocindent{3.00em}
    \coqdockw{match} \coqdocvar{res} \coqdockw{with}\coqdoceol
    \coqdocindent{3.00em}
    \ensuremath{|} \coqdocvar{LResult} (\coqdocvar{inl} \coqdocvar{xs'}) \coqdocvar{eff'{}'} \ensuremath{\Rightarrow} \coqdocvar{LResult}  (\coqdocvar{inl} (\coqdocvar{v}::\coqdocvar{xs'})) \coqdocvar{eff'{}'}\coqdoceol
    \coqdocindent{3.00em}
    \ensuremath{|} \coqdocvar{r} \ensuremath{\Rightarrow} \coqdocvar{r}\coqdoceol
    \coqdocindent{3.00em}
    \coqdockw{end}\coqdoceol
    \coqdocindent{1.00em}
    \ensuremath{|} \coqdocvar{Result} (\coqdocvar{inr} \coqdocvar{ex}) \coqdocvar{eff'} \ensuremath{\Rightarrow} \coqdocvar{LResult} (\coqdocvar{inr} \coqdocvar{ex}) \coqdocvar{eff'}\coqdoceol
    \coqdocindent{1.00em}
    \ensuremath{|} \coqdocvar{Failure} \ensuremath{\Rightarrow} \coqdocvar{LFailure}\coqdoceol
    \coqdocindent{1.00em}
    \ensuremath{|} \coqdocvar{Timeout} \ensuremath{\Rightarrow} \coqdocvar{LTimeout}\coqdoceol
    \coqdocindent{1.00em}
    \coqdockw{end}\coqdoceol
    \coqdocnoindent
    \coqdockw{end}.
    
    \caption{The functional big-step semantics of list evaluation}
    \label{Fig:functional_list_semantics}
\end{figure}

\begin{figure}[t!]
    \coqdockw{Fixpoint} \coqdocvar{eval\_fbos\_expr} (\coqdocvar{clock} : \coqdocvar{nat}) ($\Gamma$ : \coqdocvar{Environment}) (\coqdocvar{exp} : \coqdocvar{Expression})\coqdoceol
    \coqdocindent{0.50em}(\coqdocvar{eff} : \coqdocvar{SideEffectList}) \{\coqdockw{struct} \coqdocvar{clock}\} : \coqdocvar{ResultType} :=\coqdoceol
    \coqdocnoindent
    \coqdockw{match} \coqdocvar{clock} \coqdockw{with}\coqdoceol
    \coqdocnoindent
    \ensuremath{|} 0 \ensuremath{\Rightarrow} \coqdocvar{Timeout}\coqdoceol
    \coqdocnoindent
    \ensuremath{|} \coqdocvar{S} \coqdocvar{clock'} \ensuremath{\Rightarrow}\coqdoceol
    \coqdocindent{1.50em}
    \coqdockw{match} \coqdocvar{exp} \coqdockw{with}\coqdoceol
    \coqdocindent{1.50em}
    \ensuremath{|} \coqdocvar{ELit} \coqdocvar{l} \ensuremath{\Rightarrow} \coqdocvar{Result} (\coqdocvar{inl} (\coqdocvar{VLit} \coqdocvar{l})) \coqdocvar{eff}\coqdoceol
    \coqdocindent{1.50em}
    \ensuremath{|} \coqdocvar{EVar} \coqdocvar{v} \ensuremath{\Rightarrow} \coqdocvar{Result} (\coqdocvar{get\_value} $\Gamma$ (\coqdocvar{inl} \coqdocvar{v})) \coqdocvar{eff}\coqdoceol
    \coqdocindent{1.50em}
    \ensuremath{|} \coqdocvar{EFunId} \coqdocvar{f} \ensuremath{\Rightarrow} \coqdocvar{Result} (\coqdocvar{get\_value} $\Gamma$ (\coqdocvar{inr} \coqdocvar{f})) \coqdocvar{eff}\coqdoceol
    \coqdocindent{1.50em}
    \ensuremath{|} \coqdocvar{EFun} \coqdocvar{vl} \coqdocvar{e} \ensuremath{\Rightarrow} (\coqdocvar{inl} (\coqdocvar{VClos} $\Gamma$ [] \coqdocvar{vl} \coqdocvar{e})) \coqdocvar{eff}\coqdoceol
    \coqdocindent{1.50em}
    \ensuremath{|} \coqdocvar{EApp} \coqdocvar{exp} \coqdocvar{l} \ensuremath{\Rightarrow}\coqdoceol
    \coqdocindent{3.00em}
    \coqdockw{match} \coqdocvar{eval\_fbos\_expr} \coqdocvar{clock'} $\Gamma$ \coqdocvar{exp} \coqdocvar{eff} \coqdockw{with}\coqdoceol
    \coqdocindent{3.00em}
    \ensuremath{|} \coqdocvar{Result} (\coqdocvar{inl} \coqdocvar{v}) \coqdocvar{eff'} \ensuremath{\Rightarrow} \coqdoceol
    \coqdocindent{4.00em} 
    \coqdockw{let} \coqdocvar{res} := \coqdocvar{eval\_elems} (\coqdocvar{eval\_fbos\_expr} \coqdocvar{clock'}) $\Gamma$ \coqdocvar{l} \coqdocvar{eff'} \coqdoctac{in}\coqdoceol
    \coqdocindent{4.50em}
    \coqdockw{match} \coqdocvar{res} \coqdockw{with}\coqdoceol
    \coqdocindent{4.50em}
    \ensuremath{|} \coqdocvar{LResult} (\coqdocvar{inl} \coqdocvar{vl}) \coqdocvar{eff'{}'} \ensuremath{\Rightarrow}\coqdoceol
    \coqdocindent{5.50em}
    \coqdockw{match} \coqdocvar{v} \coqdockw{with}\coqdoceol
    \coqdocindent{5.50em}
    \ensuremath{|} \coqdocvar{VClos} \coqdocvar{ref} \coqdocvar{ext} \coqdocvar{varl} \coqdocvar{body} \ensuremath{\Rightarrow} \coqdoceol\coqdocindent{7.50em}
    \coqdockw{if} \coqdocvar{Nat.eqb} (\coqdocvar{length} \coqdocvar{varl}) (\coqdocvar{length} \coqdocvar{vl})\coqdoceol
    \coqdocindent{7.50em}
    \coqdockw{then}
    \coqdocvar{eval\_fbos\_expr} \coqdocvar{clock'} (\coqdocvar{append\_vars\_to\_env} \coqdocvar{varl} \coqdocvar{vl} (\coqdocvar{get\_env} \coqdocvar{ref} \coqdocvar{ext}))
    \coqdoceol
    \coqdocindent{16.60em}\coqdocvar{body} \coqdocvar{eff'{}'}\coqdoceol
    \coqdocindent{7.50em}
    \coqdockw{else} \coqdocvar{Result} (\coqdocvar{inr} (\coqdocvar{badarity} \coqdocvar{v})) \coqdocvar{eff'{}'}\coqdoceol
    \coqdocindent{5.50em}
    \ensuremath{|} \coqdocvar{\_}                             \ensuremath{\Rightarrow} \coqdocvar{Result} (\coqdocvar{inr} (\coqdocvar{badfun} \coqdocvar{v})) \coqdocvar{eff'{}'}\coqdoceol
    \coqdocindent{5.50em}
    \coqdockw{end}\coqdoceol
    \coqdocindent{4.50em}
    \ensuremath{|} \coqdocvar{LResult} (\coqdocvar{inr} \coqdocvar{ex}) \coqdocvar{eff'{}'} \ensuremath{\Rightarrow} \coqdocvar{Result} (\coqdocvar{inr} \coqdocvar{ex}) \coqdocvar{eff'{}'}\coqdoceol
    \coqdocindent{4.50em}
    \ensuremath{|} \coqdocvar{LFailure} \ensuremath{\Rightarrow} \coqdocvar{Failure}\coqdoceol
    \coqdocindent{4.50em}
    \ensuremath{|} \coqdocvar{LTimeout} \ensuremath{\Rightarrow} \coqdocvar{Timeout}\coqdoceol
    \coqdocindent{4.50em}
    \coqdockw{end}\coqdoceol
    \coqdocindent{3.50em}
    \ensuremath{|} \coqdocvar{r} \ensuremath{\Rightarrow} \coqdocvar{r}\coqdoceol
    \coqdocindent{3.50em}
    \coqdockw{end}\coqdoceol
    \coqdocindent{1.50em}
    $\cdots$\coqdoceol
    \coqdocindent{1.50em}
    \coqdockw{end}\coqdoceol
    \coqdockw{end}.
    
    \caption{Functional big-step semantics of our case study language}
    \label{Fig:functional_big_step}
\end{figure}

As we have seen before, we have to define the semantics for lists of expressions too. For these lists, we can define the functional big-step semantics distinctly, just like in case of the other discussed semantics (see the \textit{eval\_all}, \textit{eval\_prefix} predicates in Section \ref{Sec:inductive} and the \coqdocvar{AList} constructor, \ref{OS:pretty_list_step}, \ref{OS:pretty_list_exc} and \ref{OS:pretty_list_fin} in Section \ref{Sec:pretty}). So we define also a result type for list evaluation (Figure \ref{Fig:functional_values}).

Now, we have the result types, we can define the functional semantics (Figure \ref{Fig:functional_big_step} shows a representative part of it). The first step of this function is to check whether the clock is already consumed, in this case, the function returns the \coqdocvar{Timeout} value. Otherwise, the expression evaluation can begin. For calls and applications, we need the above-mentioned list evaluation. This problem is solved by the other semantics function (see Figure \ref{Fig:functional_list_semantics}), where we pass the curried version of the original functional big-step semantics as an argument (we decrease clock only in the original functional big-step semantics, so that Coq can find the decreasing argument of the function to ensure termination, moreover it enables us to use simple, yet powerful induction over the clock). Note that we decided to decrease the clock value on every nested recursive call, otherwise Coq cannot find the decreasing argument of the semantics function (however, this problem could be solved with the \verb|Program| construction possibly too). We also note, that exceptions, failures and timeouts were be handled together (except in the semantics of \coqdocvar{ETry}), because these results just needed to be propagated resulting in a short semantics definition.

\section{Discussion}\label{Sec:discussion}

In this section we evaluate and compare the semantics definitions given in the previous sections, and we also discuss a coinductive approach to handle divergence.

\subsection{Evaluation}

First of all, we can notice that all the three semantics can handle the evaluation of list of expressions separately from the body of the semantics: \textit{eval\_all} and \textit{eval\_prefix} in the traditional big-step, \ref{OS:pretty_list_exc}, \ref{OS:pretty_list_fin}, and \ref{OS:pretty_list_step} in the pretty-big-step, and \coqdocvar{eval\_elems} in the functional big-step semantics.

In terms of definition size and complexity, the functional big-step definition is superior, it is much more compact than the other two. Besides, the pretty-big-step definition uses more inference rules than the traditional big-step semantics, but these rules are much simpler (they have at most two premises). This difference increases the number of subgoals in proofs in case of the pretty-big-step semantics, but these goals are usually simpler than in the other case. In turn, simpler goals could mean simpler proofs, but since the pretty-big-step semantics is defined by mutually inductive types, the related proofs in some cases can become rather complex due to involving mutual induction.

\paragraph{Expression evaluation.}

\ifpgfplots

\begin{figure}[ht!]
	
\begin{subfigure}[ht!]{0.49\textwidth}
\includegraphics[width=7cm]{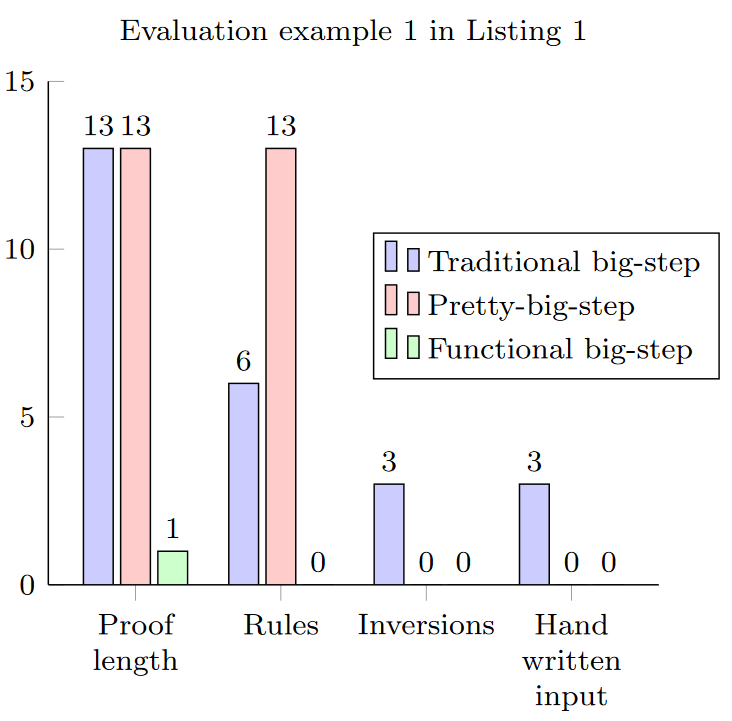}
\end{subfigure}
\begin{subfigure}[ht!]{0.5\textwidth}
    \includegraphics[width=7cm]{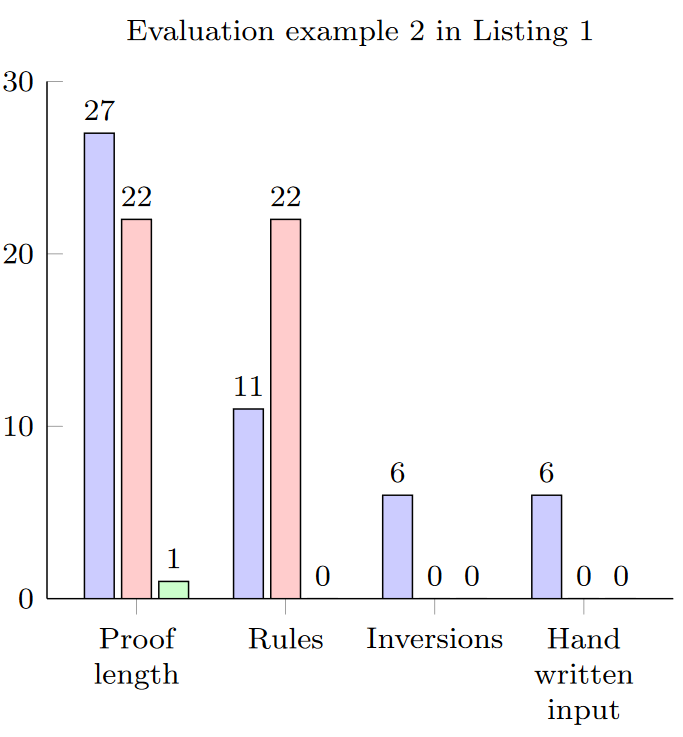}
\end{subfigure}
\begin{subfigure}[ht!]{0.49\textwidth}
\includegraphics[width=7cm]{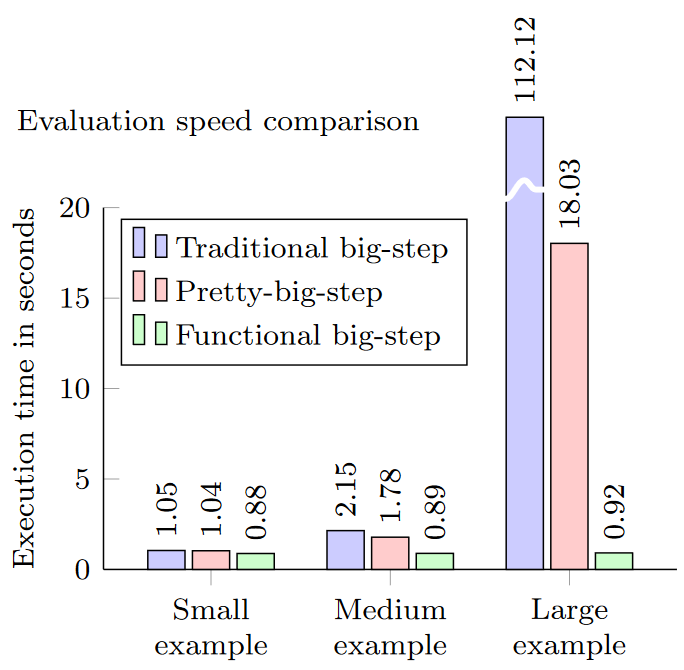}
\end{subfigure}
\begin{subfigure}[ht!]{0.5\textwidth}
    \includegraphics[width=7cm]{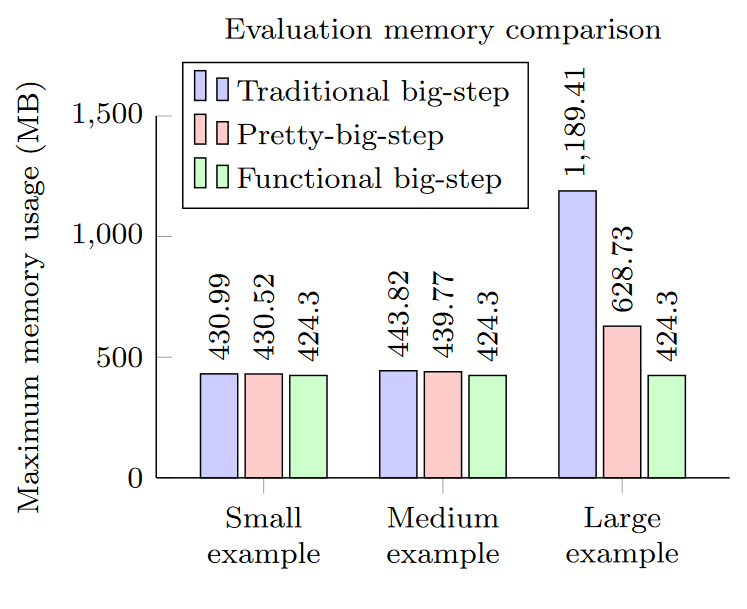}
\end{subfigure}
	\caption{Comparison of formal expression evaluation}
	\label{fig:diagram1}
\end{figure}

\fi

The first problem we encounter is that the traditional big-step and pretty-big-step semantics are relational semantics (defined by an inductive type) and are not inherently executable. To describe an expression evaluation, we need to prove the evaluation step-by-step using the inference rules (the constructors of the inductive type). As discussed before, we can also create an evaluation tactic, which can find the proof for expression evaluation, however, the use of this tactic is not efficient: it takes unreasonable amounts of memory and time (see Figure \ref{fig:diagram1}). The use of pretty-big-step semantics is more efficient than the traditional one, because for one goal, one derivation rule can match syntactically at most (except in case of different application exceptions), thus no backtracking is needed. However, it is still not efficient enough, especially compared to the functional approach.

We can also see (Figure \ref{fig:diagram1}) that the proof length of simple expression evaluations in the traditional and pretty-big-step semantics is similar. However, in the pretty-big-step semantics we used much more inference rules to reach the result, while with the traditional semantics, we had to use inversion tactic several times and specify results by hand. This is because expression list evaluation is not described step-by-step, but in universally quantified predicates, we needed to input the result list of values and side effects (e.g. \coqdocvar{eff} and \coqdocvar{vs} in \textit{eval\_all}) during formal evaluation (alternatively, list unfolding lemmas\footnote{For example a list of length $n$ can be described as $[x_1, x_2, .., x_n]$ with the $x_1, x_2, .., x_n$ existential variables.} based on the length can also solve this problem). This issue is not present in the pretty-big-step semantics, because lists are handled in a step-by-step way by \ref{OS:pretty_list_step} (again, resembling small-step evaluation). All in all, the complexity of these proofs are similar in both relational approaches.

On the other hand, the functional big-step semantics is inherently executable (because it is just a recursive function), so expressions can be simply evaluated using it, we just have to pick an appropriate initial clock value (recursion limit).

\paragraph{Expression equivalence proofs.}

\ifpgfplots

\begin{figure}[ht!]
	
	\begin{subfigure}[ht!]{0.5\textwidth}
\includegraphics[width=7cm]{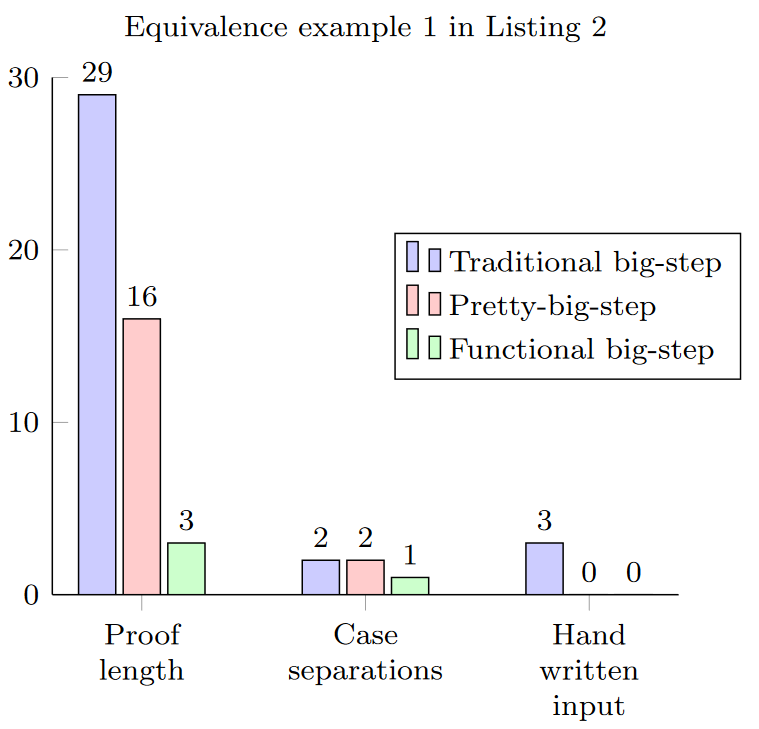}
	\end{subfigure}
	\begin{subfigure}[ht!]{0.5\textwidth}
        \includegraphics[width=7cm]{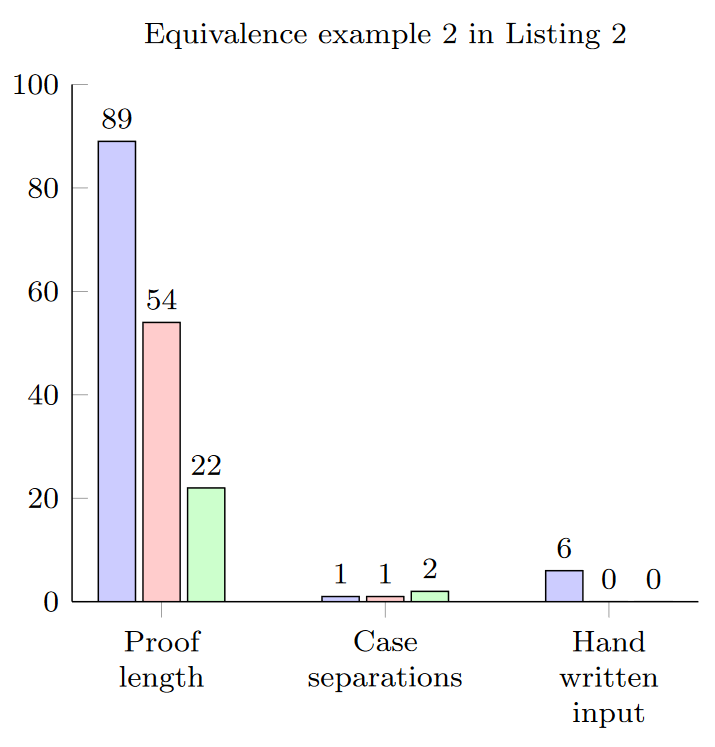}
	\end{subfigure}
	
	\caption{Complexity of expression equivalence proofs}
	\label{fig:diagram2}
\end{figure}

\fi

As we can see (Figure \ref{fig:diagram2}), surprisingly the expression equivalence proofs were the most complex in the traditional big-step semantics, while the functional big-step style performed very well. This is because in the traditional semantics we had to use list unfolding lemmas several times, which quickly increased the size of the proofs.

The use of pretty-big-step semantics was quite straightforward, and the equivalence proofs were not too complex. Once again, this is partly because no list unfolding lemmas were needed.

Based on the diagrams, one could think that there is no disadvantage of using functional big-step semantics, but that is not the case. While interactively proving the equivalences (and also semantics properties), the intermediate subgoals and assumptions were hard to read and understand, because Coq usually oversimplified the function definition, and we often saw the whole definition of this semantics, and not just necessary parts of it.
We used \verb|remember| tactics~\cite{coqltac} on the clock values which prevented the oversimplification, however, this solution is not the most convenient one.

\paragraph{Semantics property proofs.}

\ifpgfplots

\begin{figure}[ht!]
    \includegraphics[width=14cm]{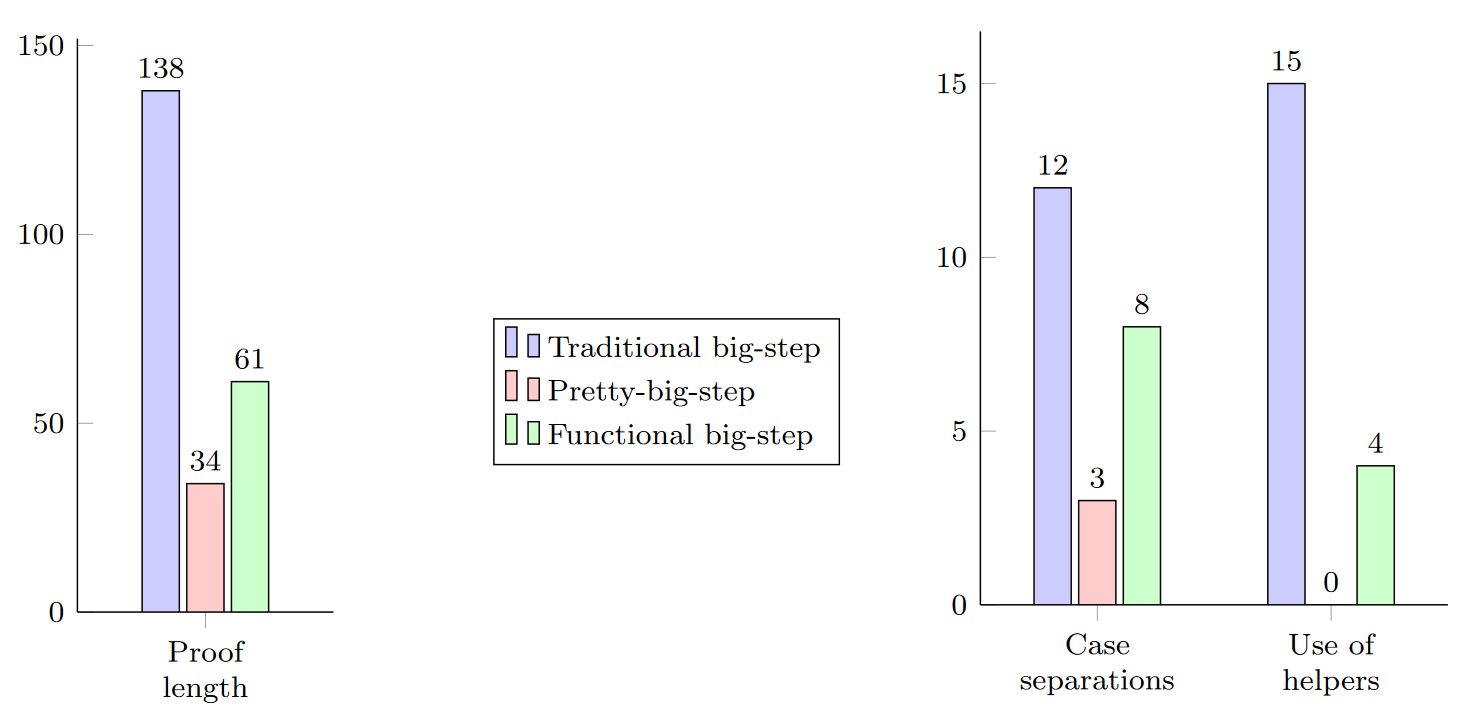}
	\centering
	\caption{Complexity of semantics property proofs}
	\label{fig:diagram3}
\end{figure}

\fi

For semantics property proofs, we chose determinism in case of the pretty-big-step and traditional big-step semantics, while a clock increasing lemma in case of the functional big-step semantics\footnote{When we evaluate an expression and get a result with the constructor \coqdocvar{Result}, then we can increase the initial clock, and get the same result.}. The determinism proof for the traditional approach is very complex. We needed to use various helper theorems about (partial) evaluation of lists of expressions and a lot of case distinctions. Still, the proof is quite long.

Proving determinism of the pretty-big-step approach was very simple, we did not need to create any helper lemmas, we used only a few case distinctions and the proof is short in spite of having to use mutual induction principle.

The proof complexity of the clock increasing lemma for the functional big-step semantics is between the previous two. We had to create and prove one helper theorem, and use several case distinctions. However, this proof is not as complex as the determinism of the traditional style, we have used simple induction over the clock variable. This style of induction is usable, because the clock is decreased in every recursive call of the semantics\footnote{Alternatively, we could have used functional induction (similar to the one mentioned by Owens \emph{et al.}~\cite{owens_functional}), however, we faced the limitations of Coq when trying to generate the induction principle.}. We should also note that while interactively proving this theorem, the subgoals were difficult to understand because of the reasons mentioned before.

In addition, we also proved the equivalence of these approaches: between pretty-big-step and functional approaches, and between traditional big-step and functional approaches by induction. Thereafter using the previous two, we also proved the equivalence of traditional and pretty-big-step semantics. We encountered one difficulty: while proving the equivalence of pretty and functional big-step semantics, the mutual induction could not be used (in functional big-step semantics, we can not give a meaning for ``intermediate terms''). To solve this problem, we followed the footsteps of Charguéraud's~\cite{pretty_big_step_formalisation} formalisation, and defined another (equivalent) version of the pretty-big-step semantics, equipped with a counter which increased when using the derivation rules of the semantics, in order to use induction over this counter. We proved the equivalence using this semantics as an intermediate step.

\subsection{Coinductive Approach}

There are two concept which were not investigated in detail: concurrency and divergence. In general, a big-step semantics can not express concurrency efficiently, because it can not handle interleaving. For this purpose, a small-step approach is more suitable.

We should note that functional big-step semantics can handle divergence in the same way Owens \emph{et al.}~\cite{owens_functional} described: the evaluation is divergent, when for any possible clock value the result is \coqdocvar{Timeout}. We also proved an expression evaluation divergent using this idea and an induction on the clock.

The previously described relational approaches are suitable to describe semantics of terminating expressions, however, they can not effectively express divergence. If one wants to reason about divergence too, a coinductive big-step semantics can be used. We have found the work of Leroy and Grall~\cite{leroy2008coinductive} the most influential, where they define a semantics for $\lambda$-calculus extended with constants. They also extend this semantics with traces, a similar feature to our side effect logging approach. Moreover, they also implemented these semantics in Coq and the source is available publicly.

We followed their footsteps to define a coinductive big-step semantics for our case-study language (in particular, for applications) with a distinct relation. For the divergence rules, we needed infinite traces for side effects too. However, this approach is not straightforward to use because of the guardedness of subgoals and we are still investigating this issue.

\section{Conclusion and Future Work}\label{Sec:conclusion}

In conclusion, we defined various approaches (primarily traditional big-step, pretty-big-step and functional big-step semantics) to define the semantics of a functional programming language, and used a small subset of sequential Core Erlang as a case study. We proved the equivalence of these semantics and evaluated, compared them from different aspects in order to choose the most fitting way to reason about refactoring correctness. Every one of these has its advantages and disadvantages. Our main three aspects were the executability of the approach, the complexity of expression equivalence proofs and proofs about the properties of the semantics, and from this point of view, the functional big-step style semantics proved to be the most useful. We highlight the fact, that these semantics and proofs are all formalised in Coq~\cite{semantics_comp}.

In the future, we are planning to formalise functional big-step semantics for sequential Core Erlang to enable effective testing of the semantics and then use comparative testing of our Core Erlang semantics and a small-step Erlang semantics~\cite{kerl} defined by one of our former project members. Naturally, we also plan to prove the existing big-step and the mentioned functional big-step semantics equivalent, after having finished the implementation. We also plan to investigate the coinductive approach more in detail. Our long term goal is to formalise entire Core Erlang and Erlang in Coq to reason about refactoring correctness on Erlang programs.

\section*{Acknowledgements}

This work was supported by the project ``Integrált kutatói utánpótlás-képzési program az informatika és számítástudomány diszciplináris területein (Integrated program for training new generation of researchers in the disciplinary fields of computer science)'', No. EFOP-3.6.3-VEKOP-16-2017-00002. The project has been supported by the European Union and co-funded by the European Social Fund.

``Application Domain Specific Highly Reliable IT Solutions'' project has been implemented with the support provided from the National Research, Development and Innovation Fund of Hungary, financed under the Thematic Excellence Programme TKP2020-NKA-06 (National Challenges Subprogramme) funding scheme.

\bibliographystyle{plain}
\bibliography{bibliography}
\end{document}